\documentclass[pra,showpacs,twocolumn,superscriptaddress]{revtex4}
\usepackage{amsmath}
\usepackage{amsfonts}
\usepackage{graphicx}

\begin{document}
\title{Continuous-variable quantum teleportation of entanglement}

\author{Tyler J.\ Johnson}
\affiliation{California Institute of Technology, Pasadena, California 91125}
\affiliation{Department of Physics and Centre for Advanced
        Computing -- Algorithms and Cryptography,       \\
        Macquarie University, Sydney,
        New South Wales 2109, Australia}
\author{Stephen D.\ Bartlett}
\author{Barry C.\ Sanders}
\affiliation{Department of Physics and Centre for Advanced
        Computing -- Algorithms and Cryptography,       \\
        Macquarie University, Sydney,
        New South Wales 2109, Australia}

\date{August 15, 2002}

\begin{abstract}
  Entangled coherent states can be used to determine the entanglement
  fidelity for a device that is designed to teleport coherent states.
  This entanglement fidelity is universal, in that the calculation is
  independent of the use of entangled coherent states and applies
  generally to the teleportation of entanglement using coherent
  states.  The average fidelity is shown to be a poor indicator of the
  capability of teleporting entanglement; i.e., very high average
  fidelity for the quantum teleportation apparatus can still result in
  low entanglement fidelity for one mode of the two-mode entangled
  coherent state.
\end{abstract}
\maketitle

\section{Introduction}
\label{sec:introduction}

Quantum teleportation~\cite{Ben93}, whereby a quantum state of a
system can be transferred by a sender Alice to a remote receiver Bob
through the use of classical communication and a shared entanglement
resource, is a remarkable demonstration of how non-local correlations
in quantum mechanics can be used to advantage.  More than simply a
novelty, quantum teleportation is useful for quantum information
processing; for example, it can be used as a universal primitive for
quantum computation~\cite{Got99}, as a fundamental component to
quantum computation with linear optics~\cite{KLM01}, and as a means to
implement non-local quantum transformations~\cite{Eis00}.  One
realization is the quantum teleportation of continuous variables
(CV)~\cite{Bra98}, which teleports states of dynamical variables with
continuous spectra; such a realization allows for the teleportation of
quantum states of light.

Any realistic (imperfect) quantum teleportation device can be
characterized by a figure of merit to quantify its ability to perform
successful teleportation.  The \emph{fidelity} $F$ is useful as a
measure of the distinguishability of the output state from the input
state, although the threshold for demonstrating genuine quantum
teleportation is debatable.  One such threshold demonstrates that an
entanglement resource was used during the experiment~\cite{Bra98}.
Another threshold~\cite{Ben93,Gro01} is that Alice teleports the state
to Bob without learning about the state in question.  For
teleportation of a distribution of states in a set~$\mathcal{C}$, one
can define the average fidelity~$\bar{F}$ of the device as the average
of the fidelity $F$ over $\mathcal{C}$.  The average fidelity provides
a figure of merit for the device to teleport states in this set.

However, any quality measure for a useful quantum teleportation
apparatus must be based on its intended application.  One important
application is that the device allows for the teleportation of
entanglement, i.e., that entanglement is preserved between the
teleported system and another (unteleported) one~\cite{Tan99}.
Consider a user, Victor, who acts to verify that the device functions
as advertised.  Alice and Bob claim that their device teleports
coherent states with a particular average fidelity, as in experimental
quantum teleportation~\cite{Fur98}.  Victor wishes to test the ability
of this device to teleport entanglement, but is restricted to
supplying Alice with coherent states according to the advertised
capability (that is, Alice must receive states in the specified set
$\mathcal{C}$).

One option is for Victor to employ a two-mode entangled coherent state
(ECS)~\cite{San92}.  By supplying Alice with only one mode, Victor can
conceal from her and Bob that they are teleporting only one portion of
a two-mode entangled state.  Bob returns the state to Victor so that
he can test whether the ECS is reconstructed with high fidelity or
not, and thus whether the teleportation has preserved the
entanglement.  Alice and Bob specified that the advertised $\bar{F}$
applies only to coherent states.  However, if they decide to check,
they will indeed see that they are receiving a mixture of coherent
states from Victor, so the supply of states from Victor does not
violate the specification that these states are drawn from a
distribution of coherent states.  Thus, Victor uses these ECSs to
quantify the capability of this quantum teleportation to preserve
entanglement.  Provided that these states are entanglements of
coherent states that are nearly orthogonal, Alice and Bob cannot
detect that Victor is using ECSs to verify the efficacy of the scheme.

We show in this paper that ECSs are useful for testing the ability of
a device to teleport entanglement.  Moreover, we show that this
entanglement fidelity does not depend on using these ECSs, but applies
generally to the teleportation of entanglement for a device that
teleports coherent states.  We note that quantum teleportation of ECSs
has been studied, but in entirely different contexts.  One such
investigation is the teleportation of ECSs in their
entirety~\cite{Wan01}, and another consideration has been to use an
ECS as a substitute for the standard entanglement resource provided by
the two-mode squeezed vacuum state~\cite{Enk01c}.  Our study is quite
different from these two cases; in our investigation, Victor employs
ECSs to replicate the conditions that Alice and Bob experience in the
experiment of Furusawa \emph{et al}~\cite{Fur98}, and Victor uses a
second entangled mode to verify that quantum teleportation of
entanglement is taking place.

The paper is constructed as follows. In Sec.~\ref{sec:quantum}, we
develop the theory of quantum teleportation of coherent states
according to a formalism that is useful for subsequent sections. In
Sec.~\ref{sec:ECS}, we discuss the quantum teleportation using ECSs as
a means of verifying the capability of teleporting entanglement; we
include a discussion of the entanglement fidelity as a measure of this
capability.  We define a noisy quantum teleportation scheme in
Sec.~\ref{sec:noisy}, and present the result that the entanglement
fidelity for the noisy quantum teleportation of ECSs is extremely
sensitive to very small errors in Alice's measurement.  We conclude
with Sec.~\ref{sec:conc}.

\section{Quantum teleportation of coherent states}
\label{sec:quantum}

Quantum teleportation was proposed~\cite{Ben93} as a means by which a
quantum state can be transferred from a system A (Alice) to a remote
system B (Bob) by employing only classical communication and a shared
entanglement resource.  Let Alice hold an arbitrary quantum state
$|\psi\rangle_1 \in \mathcal{H}_1$ that she wishes to send to Bob.  In
CV quantum teleportation, $\mathcal{H}_1$ is an infinite dimensional
Hilbert space; typically, the Hilbert space for a harmonic oscillator
is used. In addition to this quantum system, Alice holds a second
quantum system with Hilbert space $\mathcal{H}_2$, and Bob holds a
third with Hilbert space $\mathcal{H}_3$.  These two systems are in
the two-mode squeezed state~\cite{Enk99}
\begin{equation}
  \label{eq:EPRstate}
  |\eta\rangle_{23} = \sqrt{1-\eta^2}
  \sum_{n=0}^\infty \eta^n |n\rangle_2 |n\rangle_3 \, ,
\end{equation}
with $|n\rangle$ the $n$-boson Fock state.  In the limit that $\eta
\to 1$, Alice and Bob share a maximally entangled
Einstein-Podolsky-Rosen (EPR) state.

To perform quantum teleportation of the unknown state
$|\psi\rangle_1$, Alice begins by performing a joint projective
measurement on $\mathcal{H}_1 \otimes \mathcal{H}_2$ of the form
\begin{equation}
  \label{eq:idealPOVM}
  \Pi_\alpha = D_1(\alpha) |\eta\rangle_{12}\langle\eta|
  D^\dag_1(\alpha)\, ,
\end{equation}
where $D_1(\alpha)$ is the displacement operator on system~$1$ defined by
\begin{equation}
  \label{eq:BobDisplacement}
  D_1(\alpha) = \exp( \alpha \hat{a}_1^\dag - \alpha^* \hat{a}_1 ) \,
  , \quad \alpha \in \mathbb{C} \, .
\end{equation}
A measurement result is a complex number $\alpha_0 = x_0 + i p_0$.
Such a measurement can be implemented by mixing the two states on a
beamsplitter and performing balanced homodyne detection on each of the
two output modes~\cite{Bra98}.  Alice sends this measurement result
via a classical channel to Bob, who performs a displacement operation
$D_3(\alpha_0)$ (realizable by mixing with a strong coherent field
with adjustable amplitude and phase) on his system $\mathcal{H}_3$.
Bob's system is now in the state $|\psi\rangle_3$, which is identical
to the initial state $|\psi\rangle_1$ received by Alice in the limit
$\eta \to 1$.

In an experiment~\cite{Fur98}, various constraints (including finite
squeezing and imperfect projective measurements) limit the performance
of the quantum teleportation.  One must then define a figure of merit
to describe how well a physical quantum teleportation device
approximates the ideal case.  One such measure is the \emph{average
  fidelity} of the process.  Consider a pure input state
$|\psi\rangle_1$, which is imperfectly teleported such that Bob
receives the generally mixed state $\rho_3$.  The fidelity of the
output state compared to the input state is given by
\begin{equation}
  \label{eq:Fidelity}
  F(|\psi\rangle_1, \rho_3) =
  {}_1\langle\psi|\rho_3|\psi\rangle_1 \, . 
\end{equation}
(Note that some authors define the fidelity to be the square root of
this quantity.)

For a given distribution $\mathcal{C}$ of input states to be
teleported, the average fidelity $\bar{F}$ is defined to be the
weighted average of the fidelity over $\mathcal{C}$.  In the
experiment by Furusawa \emph{et al}~\cite{Fur98}, a distribution of
coherent states with fixed amplitude varied over all phases was chosen
to test the teleportation.  It is also possible to employ a
distribution over both amplitude and phase~\cite{Bow02}.

\section{Quantum teleportation of entangled coherent states}
\label{sec:ECS}

The average fidelity serves well as a figure of merit for certain
applications.  However, any measure quantifying the performance of a
quantum teleportation device must be placed in the context of its
intended use.  In particular, one may ask how well a device performs
at the important task of teleporting entanglement.

Victor, who wishes to test Alice and Bob's quantum teleportation
device, supplies Alice with a (possibly mixed) quantum state $\rho$,
and after the teleportation, Bob returns a state $\rho'$ to Victor.
Victor can then perform measurements on $\rho'$ to determine the
success or failure of the quantum teleportation.

In offering their services, Alice and Bob can be expected to quote
Victor a measure of performance of their quantum teleportation device
(such as the average fidelity), as well as a restriction on the type
of states that they can quantum teleport.  For example, they may
advertise an average fidelity $\bar{F}$ for quantum teleportation
of coherent states sampled from some distribution.  Depending on his
intended use of the quantum teleportation device, these particular
measures of performance may not be adequate.

\subsection{Entangled coherent states}

For example, Victor may wish to teleport one component of an entangled
state, and ensure that the final state returned by Bob is still
entangled with the system he kept.  If Alice and Bob advertise that
they can only teleport distributions of coherent states, it is
important that the state $\rho$ that Victor supplies to Alice is
indeed in the allowed set.  Consider the two-mode ECS~\cite{San92}
\begin{equation}
  \label{eq:ECS}
  |\Psi(\alpha,\beta)\rangle_{ab} = N\bigl(|\alpha\rangle_a |\beta\rangle_b
   - |\beta\rangle_a |\alpha\rangle_b \bigr) \, , \quad \alpha \neq
   \beta \, ,
\end{equation}
with $N = (2-2\exp(-|\alpha-\beta|^2))^{-1/2}$ the normalization.
This state is not separable, and thus possesses entanglement between
modes $a$ and $b$~\cite{Man95}.  The reduced density matrix for mode
$b$ is
\begin{align}
  \label{eq:ReducedECS}
  \rho_b &= {\rm Tr}_1
  \bigl(|\Psi(\alpha,\beta)\rangle_{ab}\langle\Psi(\alpha,\beta)|\bigr)
  \nonumber \\
  &= N^2 \bigl( |\alpha\rangle_b\langle\alpha| +
  |\beta\rangle_b\langle\beta| - \langle \alpha|\beta\rangle
  |\alpha\rangle_b\langle \beta| - \langle \beta|\alpha\rangle
  |\beta\rangle_b\langle \alpha| \bigr) \, .
\end{align}
For the case where the overlap $\langle \alpha|\beta\rangle$ is
negligibly small, this reduced density matrix is indistinguishable
from 
\begin{equation}
  \label{eq:MixedCS}
  \rho_b = \frac{1}{2}\bigl(|\alpha\rangle\langle \alpha| +
  |\beta\rangle\langle \beta|\bigr) \, ;
\end{equation}
i.e., a mixture of two coherent states.
Thus, a quantum teleportation device that functions for coherent state
inputs should function equally well for such a state.  Note that the
issues of overlaps can be avoided if Victor instead uses a two-level
system for $a$ (such as a two-level atom), and entangles coherent states
of mode $b$ to orthogonal states in $a$.  States of this form have
been experimentally realized~\cite{Bru96}.

\subsection{Entanglement fidelity}

The average fidelity may not be a good indicator of the quality of
quantum teleportation.  In particular, it overstates the capability to
teleport entanglement.  \emph{Entanglement
  fidelity}~\cite{Sch96,Bra98} is a superior measure to quantify the
ability of a process to preserve entanglement, and reduces to the
standard fidelity for the case of pure states (e.g., coherent states).

Consider a process (quantum channel) described by a superoperator
$\mathcal{E}$ and a generally mixed state $\rho$ as input.  Let $\rho'
= \mathcal{E}(\rho)$ be the corresponding output state.  We introduce
$|\Gamma\rangle$ as a purification of $\rho$, i.e., a pure state
obtained by introducing an ancilla system $A$ such that ${\rm
  Tr}_A(|\Gamma\rangle \langle\Gamma|) = \rho$.  The entanglement
fidelity 
\begin{equation}
  \label{eq:EntanglementFidelity}
  \mathcal{F}_e(\rho,\mathcal{E}) \equiv \langle \Gamma| (\mathcal{I}_A
  \otimes \mathcal{E}) \bigl(|\Gamma\rangle \langle \Gamma|\bigr)
  |\Gamma\rangle \, ,
\end{equation}
has many exceptional properties; see~\cite{Nie00}.  It is independent
of the choice of purification, meaning that it depends only on the
reduced density matrix $\rho_b$ supplied to Alice.  This property will
be useful in testing quantum teleportation, as Victor can choose any
purification (any choice of entanglement between the state he supplies
Alice and the state he retains) and achieve the same measure.  Also,
if $\rho = |\psi\rangle\langle\psi|$ is a pure state, the entanglement
fidelity $\mathcal{F}_e(|\psi\rangle\langle\psi|,\mathcal{E})$ reduces
to the standard fidelity $F(|\psi\rangle\langle\psi|,
\mathcal{E}(|\psi\rangle\langle\psi|))$.  Thus, the results of tests
employing the entanglement fidelity can be directly compared to the
fidelity of teleporting pure coherent states.

\section{Noisy quantum teleportation}
\label{sec:noisy}

In this section, we consider a noisy quantum teleportation device that
includes finite squeezing of the entanglement resource, propagation
errors, imperfect detectors, and other errors that introduce a
stochastic error given by a Gaussian distribution.  We show that,
whereas the average fidelity for quantum teleportation of coherent
states is quite robust against such errors, the entanglement fidelity
for the quantum teleportation of distributions of coherent states of
the form (\ref{eq:MixedCS}) drops off very rapidly even with highly
squeezed states and small errors.

In quantum teleportation, Alice must measure a complex amplitude
$\alpha_0$ via a joint measurement of the form (\ref{eq:idealPOVM})
and report this measurement result to Bob.  Bob then performs a
displacement $D(\alpha_0)$ on his state conditioned on this result.
If there is a measurement error, however, Alice may send to Bob the
values $\alpha_0 + z_i$, where $z_i$ is a complex error (an error in
both position and momentum) sampled from some ensemble $Z = \{z_i\}$
with corresponding probabilities $P=\{p(z_i)\}$.  Bob then performs
the displacement $D(\alpha_0 + z_i)$ rather than $D(\alpha_0)$.
(Equivalently, Bob's displacement operation could be subject to a
similar error.)  The result is that the quantum teleportation process
will no longer be ideal; it becomes a noisy process described by some
superoperator $\mathcal{E}$.

The probability distribution we employ is a Gaussian distribution with
variance $\sigma$, defined such that vacuum noise has a variance
$1/2$.  (Note that $\sigma$ has the units of the square of the
coherent state complex amplitude $\alpha$.)  Using a perfect
teleportation scheme (involving ideal EPR states of the form of
Eq.~(\ref{eq:EPRstate}) and ideal projective measurements given by
Eq.~(\ref{eq:idealPOVM})), but with a Gaussian-distributed error, the
teleported state $\rho'$ will be related to the input state $\rho$ by
\begin{equation}
  \label{eq:NoisyTransfer}
  \rho' = \int \frac{{\rm d}^2 z}{\pi\sigma} \exp\Bigl( -
  \frac{|z|^2}{\sigma}\Bigr) D(z)\rho D^\dag(z) \equiv
  \mathcal{E}_\sigma(\rho) \, . 
\end{equation}
One can view $\mathcal{E}_\sigma$ as the transfer superoperator for
this (noisy) process.

In~\cite{Bra98}, Braunstein and Kimble considered the effects of
finite squeezing ($\eta < 1$) and imperfect detectors.  Both of these
effects lead to Gaussian noise, described by the same superoperator as
given by Eq.~(\ref{eq:NoisyTransfer}).  Also, propagation losses can
be compensated for by linear amplification, which introduces an
associated Gaussian noise described similarly.  Thus, the variance for
the total error is given by the sum of the variances for the
individual errors as
\begin{equation}
  \label{eq:TotalError}
  \sigma = \sigma_G + \sigma_\eta + \sigma_\nu + \sigma_{\rm
  other} \, ,
\end{equation}
where $\sigma_G$ is the variance for the noise introduced by linear
amplification with gain $G$, $\sigma_\eta = \exp(-2\tanh^{-1}\eta)$
describes the effect of finite squeezing, $\sigma_\nu =
(1-\nu^2)/\nu^2$ describes the noise due to finite homodyne detection
efficiency $\nu$, and $\sigma_{\rm other}$ describes other sources of
Gaussian noise~\cite{Bra98}.  Again, all of these variances are
defined such that $\sigma = 1/2$ is the level of vacuum noise.  Thus,
an effective $\sigma$ describes the cummulative effects of a wide
variety of noise and errors in the quantum teleportation process.

In a classical picture, without employing a squeezed resource ($\eta
\to 0$), we find that $\sigma_\eta = 1$, i.e., that the output state
acquires two units of vacuum noise.  This noise is what Braunstein and
Kimble refer to as ``quantum duty'', or quduty~\cite{Bra98}: one unit
of vacuum noise is acquired by each pass across the quantum/classical
border (one by Alice's measurement and one by Bob's reconstruction).

For an input state given by a pure coherent state $|\alpha\rangle$, it
is straightforward to calculate the entanglement fidelity of the
operation.  (As the state is pure, the entanglement fidelity
$\mathcal{F}_e$ will equal the standard fidelity $F$.)  The
entanglement fidelity for the coherent state $|\alpha\rangle$,
\begin{equation}
  \label{eq:EntFofCS}
  \mathcal{F}_e(|\alpha\rangle\langle\alpha|,\mathcal{E}_\sigma) =
  \frac{1}{1+\sigma} \, ,
\end{equation}
is independent of $\alpha$.  The lower bounds on fidelity for quantum
teleportation discussed in the introduction are clear from this
equation.  The bound of $F > 1/2$ by Braunstein and
Kimble~\cite{Bra98} can only be satisfied if $\sigma < 1$; i.e., the
variance must be less than twice the vacuum noise, verifying the use
of an entangled resource.  The more stringent bound $F > 2/3$, set by
Grosshans and Grangier~\cite{Gro01} using an argument of no-cloning,
requires $\sigma < 1/2$ or up to one unit of vacuum noise.

If the state to be teleported is one mode of an ECS, the calculation
of the entanglement fidelity is more involved.  Fortunately, a
purification of the state (\ref{eq:ReducedECS}) is already provided:
it is the ECS itself.  As noted earlier, the entanglement fidelity is
independent of the choice of purification.  The entanglement fidelity
for the noisy quantum teleportation of the $b$ mode is given by
\begin{align}
  \label{eq:EntFofECS}
  \mathcal{F}_e(\rho_b,\mathcal{E}_\sigma)  
  &= \bigl\langle\Psi(\alpha,\beta)\bigl| (\mathcal{I}_a \otimes
  \mathcal{E}_{\sigma b}) \bigl(|\Psi(\alpha,\beta)\rangle \langle
  \Psi(\alpha,\beta)|\bigr) \bigr|\Psi(\alpha,\beta)\bigr\rangle 
  \nonumber \\
  &= \int \frac{{\rm d}^2 z}{\pi\sigma} \, \exp\Bigl(-
  \frac{|z|^2}{\sigma}\Bigr) \Bigl|\langle\Psi(\alpha,\beta)| D_b(z)
  |\Psi(\alpha,\beta)\rangle\Bigr|^2 \, . 
\end{align}
The expression is complicated by the nonorthogonality of coherent
states, but the overlap drops rapidly as the coherent states are
increasingly separated in phase space.  To simplify this expression,
we assume that $|\alpha-\beta|$ is sufficiently large that we can
ignore terms bounded above by $|\langle\alpha|\beta\rangle|^2 =
\exp(-|\alpha-\beta|^2)$.  Again, we note that these overlaps can be
avoided if Victor chooses to couple coherent states in $b$ to
orthogonal modes in $a$, and that this choice of purification gives
identical results for the entanglement fidelity.
Using this assumption,
\begin{widetext}
\begin{align}
  \label{eq:CalculatedEntFofECS}
  \mathcal{F}_e(\rho_b,\mathcal{E}_\sigma) &\simeq 
  N^4 \int \frac{{\rm d}^2 z}{\pi\sigma} \, \exp
  \Bigl(-\frac{|z|^2}{\sigma}\Bigr) \Bigl\{ 
  |\langle \alpha|\alpha + z\rangle|^2 + |\langle \beta|\beta+z\rangle|^2
  + 2 {\rm Re} \bigl(\exp\bigl(i {\rm
  Im}((\alpha-\beta)z^*)\bigr)\langle 
  \alpha+z|\alpha\rangle \langle \beta|\beta + z\rangle \bigr)\Bigr\} 
  \nonumber \\
  &\simeq \int \frac{{\rm d}^2 z}{2\pi\sigma}\,
  \exp\Bigl(-\frac{|z|^2}{\sigma}\Bigl) \exp(-|z|^2) \Bigl(
  1+\cos\bigl(2{\rm Im}((\alpha-\beta)z^*)\bigr)\Bigr)
  \simeq \frac{1}{2}\Bigl(\frac{1}{1+\sigma}\Bigr) \bigl(1 +
  \exp(-|\alpha - \beta|^2 \sigma_{\rm eff})\bigr) \,,
\end{align}
\end{widetext}
where $\sigma_{\rm eff} = \sigma/(1+\sigma)$.

The entanglement fidelity for the noisy quantum teleportation of one
mode of an ECS differs from that of a pure coherent
state~(Eq.~(\ref{eq:EntFofCS})) due to a term that drops exponentially
in $|\alpha-\beta|^2$.  As this term becomes negligibly small (for
even small errors described by $\sigma \ll 1$), the entanglement
fidelity for the teleportation of ECSs approaches half the value for
that of a pure coherent state.

\begin{figure}
  \includegraphics*[width=3.25in,keepaspectratio]{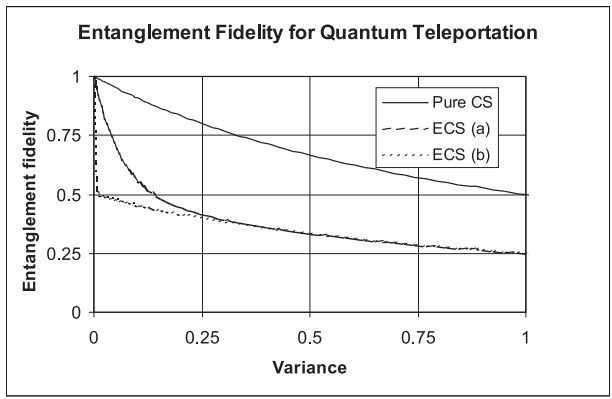}
  \caption{Entanglement fidelity as a function of the standard
    deviation $\sigma$ for the errors in noisy quantum teleportation.
    The entanglement fidelity is plotted for a pure coherent state
    (CS) $|\alpha\rangle$, and for entangled coherent states (ECS) of
    the form $|\Psi(\alpha,-\alpha)\rangle$ for (a) $\alpha = 2$ and
    (b) $\alpha = 10$.  Note that a variance of $1$ corresponds to
    two units of vacuum noise, i.e., two ``quduties''.}
  \label{fig:Fidelity}
\end{figure}

In Fig.~\ref{fig:Fidelity}, we compare the standard fidelity for
teleportation of a pure coherent state $|\alpha\rangle$ to the
entanglement fidelity for teleporting the ECS with $\beta = - \alpha$
for two values, $\alpha=2$ and $\alpha=10$.
One key feature to notice is that the entanglement fidelity for all
cases is reduced significantly for a variance on the order of the
vacuum noise $\sigma \simeq 1/2$.  Thus, the precision of quadrature
phase measurements must be very good on the scale of the standard
quantum limit~\cite{Cav81}.  For the ECSs, the rapid decrease of the
entanglement fidelity to approximately half that of the pure coherent
states is clearly evident.  Consider ECSs of the form
$|\Psi(\alpha,-\alpha)\rangle$ with mean photon number
$\bar{n}=|\alpha|^2$.  In order to maintain a constant entanglement
fidelity $\mathcal{F}_e(\rho_b,\mathcal{E}_\sigma) > 1/2$, the
variance of the errors must scale as $\sigma \sim 1/\bar{n}$.  Thus,
quantum teleportation of a single mode of an ECS with high
entanglement fidelity becomes increasingly difficult as the mean
photon number of the state is increased.

Note that, for any distribution $\mathcal{C}$ of coherent states that
leads to an average fidelity for quantum teleportation, one can also
calculate an average entanglement fidelity both for pure coherent
states and ECSs sampled from the same distribution.  In the experiment
of Furusawa~\emph{et al}, an average fidelity of $\bar{F}=0.58$ has
been obtained for the set~$\mathcal{C}=\{|\alpha_0
e^{i\varphi}\rangle;0\leq\varphi<2\pi\}$, where $|\alpha_0
e^{i\varphi}\rangle$ is the coherent state with $|\alpha_0|^2 \sim
100$ the mean photon flux and $\varphi$ is the phase of the coherent
state~\cite{Fur98,Ralph}.  This phase is uniformly distributed over
the domain $[0,2\pi)$.  The experimental result~\cite{Fur98} can be
compared against our calculation for teleportation of ECSs of the form
$|\Psi(\alpha_0,-\alpha_0)\rangle$ with $\alpha_0=10$, presented in
Fig.~\ref{fig:Fidelity}.  Whereas the entanglement fidelity for the
pure coherent state is very high for small $\sigma$, the entanglement
fidelity for the ECS is less than $0.5$ even for $\sigma = 0.01$
(corresponding to a total error around 2\% of the vacuum noise).  Note
that with otherwise perfect conditions (no propagation loss, detector
noise, etc.), at least 8.5 dB of squeezing is required to achieve an
entanglement fidelity of greater than $0.5$ for this state; this
amount of squeezing represents a target for high entanglement fidelity
quantum teleportation of distributions of coherent states.

\section{Conclusions}
\label{sec:conc}

We have shown that the average fidelity is not necessarily a good
measure of successful quantum teleportation, and in particular is
shown to be a poor indicator for the capability of quantum
teleportation to preserve entanglement.  On the other hand, the
entanglement fidelity provides a useful figure of merit for the
quantum teleportation of entanglement.  We demonstrate that
entanglement with other systems can be used to test claims of quantum
teleportation, even for a restricted set of allowed input states.  In
particular, ECSs can be used to test quantum teleportation devices
that advertise only teleportation of mixtures of coherent states.  We
show that the entanglement fidelity of distributions of coherent
states is extremely fragile, and can be drastically reduced from the
fidelity of the pure coherent states by the effects of finite
squeezing, imperfect detection, propagation errors, or small
stochastic errors in Alice's measurements (or Bob's transformations).

An important application of teleporting coherent states is in a
distributed quantum network that employs only Gaussian states and
Gaussian-preserving operations, i.e., linear optics~\cite{Bar02b}.
In such a network, the appropriate figure of merit for the
teleportation of entanglement between nodes is clearly the
entanglement fidelity.

\begin{acknowledgments}
  This project has been supported by an Australian Research Council
  Grant. TJJ has been financially supported by the California
  Institute of Technology, and appreciates the hospitality of
  Macquarie University where this project was undertaken.  SDB
  acknowledges the support of Macquarie University.  We acknowledge
  useful discussions with T.\ C.\ Ralph and T.\ Tyc.
\end{acknowledgments}

\end{document}